# HT-Paxos: High Throughput State-Machine Replication Protocol for Large Clustered Data Centers

Vinit Kumar[1] and Ajay Agarwal[2]

[1] Associate Professor with the Krishna Engineering College, Ghaziabad, India.
 (Phone: +919971087809; e-mail: vinitbaghel@gmail.com)

[2] Professor with SRM University, DELHI-NCR Campus, Modinagar, Ghaziabad, India.
 (Phone: +919917083437; e-mail: ajay.aagar@gmail.com)

Abstract

Paxos is a prominent theory of state machine replication. Recent data intensive Systems those implement state machine replication generally require high throughput. Earlier versions of Paxos as few of them are classical Paxos, fast Paxos and generalized Paxos have a major focus on fault tolerance and latency but lacking in terms of throughput and scalability. A major reason for this is the heavyweight leader. Through offloading the leader, we can further increase throughput of the system. Ring Paxos, Multi Ring Paxos and S-Paxos are few prominent attempts in this direction for clustered data centers. In this paper, we are proposing HT-Paxos, a variant of Paxos that one is the best suitable for any large clustered data center. HT-Paxos further offloads the leader very significantly and hence increases the throughput and scalability of the system. While at the same time, among high throughput state-machine replication protocols, HT-Paxos provides reasonably low latency and response time.

## 1. Introduction

State machine replication (SMR) is a fundamental technique for increasing availability of the system [1] [6]. It lies in the heart of the many real time applications. Replicating a service on multiple servers ensures that even if some replica fails the service is still available. State machine replication prevalently uses the variants of Paxos. Google's Megastore [25], chubby lock service [14] and yahoo's Zab [24] are few of the popular applications that use the variant of Paxos. Since in leader based protocols, leader does most of the work, the bottleneck is found at the leader and the maximum throughput is limited by the leader's resources (such as CPU and network bandwidth), further increasing the number of client requests by increasing more clients results in a decrease of the throughput. Since the bottleneck is at the leader, more additional replicas will not improve performance; in fact, it decreases throughput since the leader requires to process additional messages.

Ring Paxos [23] offload the leader by adopting the concepts of (i) ordering of ids by the leader instead of full requests, (ii) dissemination of requests and learned-ids by the leader through ip-multicasting, (iii) a ring of acceptors, (iv) batching of requests at leader and (v) pipelining (i.e. parallel execution of ring Paxos instances). Concept of ip-multicasting allows the leader to order the ids instead of full requests and hence offloads the leader. Ring of acceptors reduces the number of messages sent to other acceptors and received from other acceptors by the leader. Because of the ring, learners learn the decision only from the leader. In other Paxos protocols those are optimized for latency instead for messages, learners learn the decision from any quorum of acceptors i.e. Ring of acceptors reduces the load on the learners. Batching of requests at leader also significantly offload the leader. Moreover, concept of cheap Paxos reduces the latency.

However, in ring Paxos leader still requires to handle all client communications, assigns unique *id* to client requests, sends all client requests with their *id* to all acceptors and learners, forward the *id* to the first acceptor of the ring and on receive *id* from the last acceptor of the ring broadcasts their decision to all acceptors and learners. In ring Paxos, clients also require knowing about the leader; if leader fails then service will interrupt until the election of a new leader.

S-Paxos [29] offloads the leader by using the concepts as (i) distributing the work of handling all client communications among all non faulty replicas, (ii) disseminating client requests among replicas in a distributed fashion, (iii) ordering of ids by the leader instead of full client requests, (iv) batching the client requests and (v) pipelining. Receiving of client request by any replica, have certain advantages as, it offloads the leader and failure of the leader does not interrupt service.

However, in S-Paxos, every non-faulty replica including leader receives all client requests either directly from clients or through other replicas. All these client requests may reach to the leader in less number of messages because of the batching at various replicas (unlike ring Paxos). Moreover, leader may not require disseminating all client requests because of the aforementioned second concept of S-Paxos, (unlike ring Paxos) but partially disseminates the client requests and partially handles client communications. In addition, leader uses classical Paxos for ordering ids (instead of full client requests), so leader and other replicas handles the all messages belongs to classical Paxos. High number of messages at leader adversely affect throughput.

Multi-Ring Paxos [27] uses the concept of state partitioning [8] for increasing the throughput of the system. Each partition uses a different instance of ring Paxos. The performance of ring Paxos directly affects this protocol.

In this paper, we are proposing HT-Paxos (HT stands for high throughput) a variant of Paxos, that adopts all aforementioned concepts of S-Paxos for offloading the leader. In addition, HT-Paxos adopts few major concepts as (i) eliminating the work of handling client communications and request dissemination from the leader, i.e. leader does not require either receiving or disseminating the client requests instead it only receives the batch ids (or request ids) and orders them (unlike S-Paxos and ring Paxos). (ii) Significantly reducing acknowledgement messages at disseminators in large clustered data centers (unlike S-Paxos, where every disseminator sends acknowledgement messages to every other disseminator). In this way, leader as well as other disseminators becomes truly lightweight and hence for any large clustered data center, HT-Paxos provides significantly higher throughput.

Organization of this paper is as follows, Next section revisits Paxos, if you are well aware about Paxos then may skip this section. After that, Section 3 presents a system model. Moreover, Section 4 proposes the HT-Paxos. While Section 5 presents a comparative analysis of proposed work with other related work. Finally, concluding Section discusses the advantages of HT-Paxos.

## 2. Revisiting Paxos

Under this section, we are briefly reviewing the related theory of Paxos. Whereas, Paxos is a family of protocols that implements a replicated state machine and assumes a distributed system of processes communicating by messages. Processes can fail only by stopping, and messages can be lost or duplicated but not corrupted. Timely actions by non-failed processes and timely delivery of messages among them are required for progress; safety is maintained despite arbitrary delays and any number of failures.

Any of the Paxos protocol has three types of agents: *proposers*, *acceptors*, and *learners*. In an implementation, a single process may act as more than one agent. *Proposers* propose the commands. *Acceptors* choose the sequence of commands and *Learners* learn and execute the commands. If only one *Proposer* is supposed to propose all the commands and resolves the conflicts then we call it as a *Leader*.

2.1 Classical Paxos

In classical Paxos [3] [4] clients send their command to the Leader. Leader creates a separate instance of Paxos protocol for every command and assigns an instance number to each instance sequentially. If an instance of Paxos protocol at any server communicates with another server then another server creates a new instance of Paxos protocol, with the same instance number, if the instance of same instance number does not exists. Moreover, If Leader fails then leader election protocol elects a new Leader.

Every instance of Paxos protocol takes one or more rounds to decide on a single output value. Whereas a successful round has two phases:

*Phase 1a:* Proposer (Leader) selects a proposal number $n$ and sends a *Prepare* message that contains a proposal along with proposal number $n$ to a majority of Acceptors.

*Phase 1b:* If the proposal number $n$ of the current proposal is larger than the proposal number of any previous proposal, then Acceptor promises not to accept proposals less than $n$, and sends the last accepted proposal (if any) to the Proposer. Otherwise, Acceptor sends a denial to the Proposer.

*Phase 2a:* If the Proposer receives a response (numbered $n$) from a majority of Acceptors then it chooses a highest numbered proposal received from all such responses. If Proposer does not receive any accepted proposal then Proposer chooses any one of the proposed proposal. Now Proposer sends an *Accept* message to a majority of Acceptors along with a Chosen proposal and proposal number $n$.

*Phase 2b:* If an acceptor receives an *accept* message for a proposal numbered $n$ then it accepts the proposal unless it has already responded to a *prepare* message having a proposal number greater than $n$. after accepting the proposal, it sends an *Accepted* message along with accepted proposal to the Proposer and every Learner. Any round fails when multiple Proposers send conflicting *Prepare* messages or the Proposer does not receive a majority of responses. In these cases, another round starts with a higher proposal number.

In addition, different Proposers choose their proposal numbers from the disjoint sets of numbers. Therefore, two different Proposers never issue a proposal with the same proposal number. Moreover, each Proposer maintains the highest numbered proposal with proposal number in a stable storage and phase 1 always uses a higher proposal number than any it has already used. An Acceptor always records its intended response in a stable storage before actually sending the response. Furthermore, every Learner executes the learned commands sequentially as per the instance numbers.

### 2.1.1   Optimizations of classical Paxos

If leader is relatively stable then phase one becomes quite unnecessary. Thus, it is possible to skip phase one for future instances of the protocol with the same leader. To achieve this, the instance number is included along with each value. It reduces the failure-free message delay (proposal to learning) from four delays to two delays.

Another optimization reduces the number of messages, as phase 2b messages reaches only to the leader, if leader receives such messages for the same value from majority of acceptors then leader decides this value, and sends this decision to all learners. However, this optimization increases the latency.

## 2.2 Fast Paxos

Fast Paxos [5] generalizes Basic Paxos to reduce end-to-end message delays. In Basic Paxos, the message delay from client request to learning is three message delays. Fast Paxos allows two message delays, but requires the Client to send its request to multiple destinations. Intuitively, if the leader has no value to propose, then a client could send an *Accept*! message to the Acceptors directly. The Acceptors would respond as in Basic Paxos, sending *Accepted* messages to the leader and every Learner achieving two message delays from Client to Learner. If the leader detects a collision, it resolves the collision by sending Accept! Messages for a new round, which are Accepted as usual. This coordinated recovery technique requires four message delays from Client to Learner. The final optimization occurs when the leader specifies a recovery technique in advance, allowing the Acceptors to perform the collision recovery themselves. Thus, uncoordinated collision recovery can occur in three message delays (and only two message delays if all Learners are also Acceptors).

## 2.3 Generalized Paxos

Generalized Paxos [11] generalizes the classical Paxos, multi Paxos and fast Paxos. Moreover, it explores the relationship between the operations of a distributed state machine for improving performance. When conflicting proposals are commutative operations of the state machine, in such cases, coordinator accepts all such conflicting operations at once, avoiding the delays required for resolving conflicts and re-proposing the rejected operation. This Paxos uses ever-growing sets of commutative operations, after some reasonable time, these sets become stable and then leader accept this set. Larger set reduces the number of messages and time taken by the sate machine.

2.4 Ring Paxos

Ring Paxos [23] has a logical ring of acceptors. One acceptor of the ring plays a role of the coordinator (leader). Coordinator accepts client requests and assigns a unique ID to each client request. Moreover, in phase 1, Coordinator and majority of acceptors makes an agreement about the ring of acceptors. As soon as batch of client requests at coordinator completes or reaching timeout, phase two triggers. In phase 2, coordinator ip-multicasts the client requests along with their IDs, round number and instance number to all acceptors and learners. Ring Paxos executes consensus on IDs.

Upon receiving a phase 2 message, first acceptor in the ring creates a small message containing the round number, IDs and its own decision and forwards it along the logical ring. Moreover, upon receiving a message from an acceptor of the ring, other than coordinator, each acceptor in the ring appends its decision to the message and forwards it along the logical ring, if it has the corresponding client requests.

Upon receiving the phase 2 message from the last acceptor of the ring, coordinator informs all the learners that some IDs have been chosen. In high load conditions, this information can be piggybacked on the next ip-multicast message. Moreover, learner delivers the corresponding client value in the appropriate instance.

2.5 Multi-Ring Paxos

Multi-Ring Paxos [27] uses the concept of logical partitioning for increasing the throughput of the system, proposers, acceptors and learners subscribe to one or more logical partitions. Each partition uses a different instance of ring Paxos.

2.6 S-Paxos

S-Paxos [29] assumes that all the replicas servers play the roles of all agents and out of them, one replica plays a role of the leader. Moreover, any client may send their request with their unique id to any of the replica. Replica accepts client requests, and creates a batch that contains client requests and their ids. After that, replica assigns an id to that batch. Now replica forwards this batch and batch id to all the replicas including self. When a replica receives a forwarded batch with their batch id, it records the batch and the batch id in the *requests* set. It then sends an acknowledgment containing the batch id to all replicas. Replica retransmit acknowledgement message periodically until batch stabilizes. Batch stabilizes after receiving $f+1$ acknowledgments from different replicas for a particular batch id ($f$ represents here an upper bound for faulty replicas). A replica records this fact by adding the batch id to its *stableIds* set. If replica receives an acknowledgement for a particular batch id from any replica $q$, but does not has corresponding batch then it requests $q$ for resending the corresponding batch.

Moreover, the leader replica passes the batch ids available in *stableIds* set to the ordering layer, which will then use the classical Paxos protocol to order it. Here it is significant that classical Paxos achieves consensus on ids rather than full requests. Replicas execute client requests in the order as suggested by classical Paxos. After executing the request, the replica that received the request from the client sends the corresponding reply. In the low load condition, we may avoid batches but S-Paxos is designed for high throughput, therefore, batching and pipelining is quite well desirable. In high load conditions, any outgoing message that contains any batch may piggyback acknowledgement messages.

## 3. System Model

HT-Paxos is a variant of Paxos. However, we have divided the role of *acceptors* into two separate sub categories as (i) *disseminators* and (ii) *sequencers*. In this way, HT-Paxos have four classes of agents: *proposers (clients)*, *disseminators*, *sequencers* and *learners.* One *sequencer* assumes a role of the *leader*. Proposers propose Proposals (*requests*); disseminators accept proposals and disseminate accepted proposals to all other disseminators and learners; Sequencers work for establishing an order by using classical Paxos (classical Paxos uses both sequencers and learners for determining and learning an order). Learners receive proposals from disseminators and execute them in an order as indicated by the *leader*. Although agents work differently for improving throughput but fundamental guarantees (*Nontriviality, Stability, Consistency and Liveness*) of Paxos are the same in HT-Paxos.

We propose that clustered data centre have two LANs (local area networks), we call them as first LAN and second LAN. All disseminators and learners subscribe to both the LANs. Moreover, all sequencers subscribe to the second LAN. Furthermore, proposers either subscribe to the both the LANs or connect both the LANs via one or more routers.

Any computing node that has a disseminator will also have a learner and in such nodes, both agents can share all incoming messages and data structures. Moreover, nodes those have sequencers does not has any other agent. Furthermore, each computing node has two buffers one for incoming messages and another for outgoing messages for each LAN.

Like Classic Paxos, we assume that agents communicate by sending messages. These messages can take arbitrarily long for reaching their destinations, can be delivered out of order, can be duplicated, and can be lost. Moreover, system detects all corrupted messages and considers such messages (corrupts in communication medium and finally detected) as lost. Furthermore, agents discard duplicate messages, as well as learners discard duplicate proposals.

Like Classic Paxos, we assume the customary partially synchronous, distributed and non-Byzantine model of computation. Where, Agents operate at arbitrary speed, may fail by stopping, may restart and always perform an action correctly. Agents have access to stable storage whose state survives failures.

We assume that, at least $\lfloor n/2 \rfloor + 1$ disseminator will always remain non-faulty out of the total $n$ disseminators, at least $\lfloor m/2 \rfloor + 1$ sequencers will always remain non-faulty out of the total $m$ sequencers and least one learner will always be non-faulty.

Slight modifications in the system model are there for the optimized versions of HT-Paxos. We have explained these modifications in the next section.

For sending a message, we have used two primitive (i) Send < *message* > to one receiver (ii) Multicast < *message* > to multiple receivers. Send primitive is for one to one communication and Multicast primitive represents that sender sends a single message but specified multiple receivers can receive this message. We can implement this multicasting by using Ethernet/hardware multicasting, by using IP multicasting, or by using Dr. Multicast. Dr. Multicast [19] explains that IP Multicast in data centers becomes disruptive in the presence of large number of groups and requires a proper administrative control. However, in HT-Paxos we have only few groups. In addition, use of multiple LANs further reduces the number of groups per LAN.

Like S-Paxos, we divide all activities of HT-Paxos into two layers, (i) dissemination layer and (ii) ordering layer. All work performed by classical Paxos comes under ordering layer and rest of the work that one is related to the dissemination of the *request* comes under dissemination layer.

## 4. HT-Paxos

### 4.1 Basic Algorithm

#### *4.1.1 An Overview*

Any client sends their *request* (*request* contains a *request_value* and a unique *request_id*) to any one disseminator (randomly chosen) using first LAN. Moreover, if client does not receive a reply message < *request_id* > in a reasonably long time, then it periodically sends their same *request* to any one disseminator (randomly chosen) using first LAN until it gets a reply. Furthermore, if client gets a reply message, then it replies with < *request_id* > message to that disseminator using second LAN.

If *request* is available from any client then disseminator receives a *request*. After that, it multicasts this *request* using first LAN to all disseminators and learners. Moreover, when a disseminator receives a *request* from any disseminator then (i) it records the *request* in the *requests_set*, (ii) replies back an acknowledgment message < *request_id* > to that disseminator using second LAN and (iii) periodically multicasts < *request_id* > message to all sequencers using second LAN until *request_id* become an element of *decided* set.

Disseminator that received the *request* from the client sends a reply message < *request_id* > to the corresponding client using second LAN in either of the two conditions, (i) on receiving < *request_id* > message from at least a majority of disseminators (including self), or (ii) on observing that *request_id* is an element of *decided* set. Moreover, this disseminator periodically sends a reply to the corresponding client until it gets a reply message < *request_id* > or detects a failure of the client.

If disseminator does not receive sufficient desired acknowledgement messages < *request_id* > then it periodically multicasts < *request_id* > message to all disseminators using second LAN until it receives desired acknowledgment messages or when *request_id* becomes an element of *decided* set.

If any disseminator *p* receives < *request_id* > message from any disseminator *q*, but it does not has the corresponding *request*, then *p* sends a message < Resend, *request_id* > to *q* using second LAN. Moreover, on receiving < Resend, *request_id* > message from any disseminator *p*, disseminator *q* sends the corresponding *request* to the disseminator *p* using first LAN.

After receiving a < *request_id* > message from any learner, disseminator replies with the corresponding *request* to that learner.

After receiving same < *request_id* > messages from at least a majority of disseminators, sequencer inserts this *request_id* into its *stable_ids* set.

Moreover, leader repeatedly launches (up to the allowable number of instances at a time) an instance of classical Paxos for each *request_id* from the *stable_ids*. Classical Paxos uses second LAN for their all communications. After learning a *request_id,* learner inserts this *request_id* into the *decided* set.

Each disseminator also maintains *requests_set* at the permanent storage device, initially the value of this set is null and at every startup, disseminator will initialize this set through reading permanent storage device. Moreover, if learner is not at disseminator's site then learner similarly maintains this set. Furthermore, if learner is on disseminator's site then learner does not maintain this set but may read this set when required.

Each learner also maintains *decided* set at permanent storage device and initially the value of this set is null and at every startup, learner will initialize this set through reading permanent storage device. Disseminator on the same computing node may read this set when required.

Each sequencer also maintains *stable_ids* and decided sets at permanent storage device and initially the value of both these sets are null and at every startup, sequencer will initialize these sets through reading permanent storage device.

### 4.1.2 Pseudo Code of Dissemination Layer

*Algorithm 1:* Dissemination Layer of HT-Paxos

1. /* *Task of a proposer (client)* */
2. Create a new *request*
3. Choose any disseminator *d* randomly.
4. Send < *request* > to *d* using first LAN
5. **Upon** not receiving any reply message < *request id* > from any disseminator until Δ₁ time,
6.     **Repeat** from step 3
7. **Upon** receiving a reply message < *request id* > from any disseminator *d*
8.     Send < *request_id* > to *d* using second LAN
9. **If** (want to send more *requests?*)
10.     **Repeat** from step 2,
11. **Else,** exit.

12. /* *Task of a disseminator* */
13. **Upon** receiving < *request* > from any client
14.     Multicast < *request* > to all disseminators and learners using first LAN
15. **Upon** receiving < *request* > from any disseminator *d*
16.     Requests_set ← Requests_set ∪ request
17.     Send < *request_id* > to *d* using second LAN
18.     Multicast < *request_id* > to all sequencers using second LAN,

19.       **Repeat** from step 18 after every $\Delta_2$ time, **until** $\left(request\_id \in decided\right)$
20. **Upon** receiving $< request\_id >$ message from at least a majority of disseminators or $request\_id \in decided$
21.     **If** (received the corresponding *request* from the client)
22.     **Then**
23. Send $< request\_id >$ to the corresponding client using second LAN
24. **Repeat** from step 23 after every $\Delta_3$ time, **until** it receives a reply message $< request\_id >$ from the corresponding client or client's failure is detected
25. **Upon** receiving $< request\_id >$ from any disseminator $q$, $\begin{pmatrix} \forall request : request \in requests\_set \\ \land request\_id \notin request \end{pmatrix}$ and after $\Delta_4$ time
26.     Send $<$ Resend, $request\_id >$ to $q$ using second LAN
27. **Upon** receiving $<$ Resend, $request\_id >$ from any disseminator $p$
28.     Send $< request >$ to $p$ using first LAN
29. **Upon** receiving $<$Resend, $request\_id >$ from a learner $l$
30.     **If** $\begin{pmatrix} \exists request : request \in requests\_set \\ \land request\_id \in request \end{pmatrix}$
31.       Send $< request >$ to $l$ using first LAN
32. **If** $\left( request\_id \in decided \land \begin{pmatrix} request \notin requests\_set : \\ request\_id \in request \end{pmatrix} \right)$
33.     Send $<$ Resend, $request\_id >$ to any other disseminator using second LAN
34.     **Upon** not receiving corresponding *request* after $\Delta_5$ time **Repeat** from step 32

35. /* *Task of a sequencer* */
36. **Upon** receiving same $< request\_id >$ from at least a majority of disseminators
37.     *Stable_ids* ← *Stable_ids* ∪ *request_id*

38. /* *Task of a learner* */
39. **If** (learner is not at disseminator's site)
40.     **Then**
41.     **Upon** receiving $< request >$ from any disseminator
42.     *Requests_set* ← *Requests_set* ∪ *request*
43.     **If**
$$\left( \begin{pmatrix} \forall request\_id : \\ \text{learner has learned } request\_id \land \\ \left( request \notin requests\_set : request\_id \in request \right) \end{pmatrix} \right)$$
44.     Send $<$Resend, $request\_id >$ to any disseminator using second LAN
45.     **Upon** not receiving corresponding *request* after $\Delta_6$ time **Repeat** from step 43
46. **Execute** *requests* in an order as provided by ordering layer.

─────────────────────────────────────────

### 4.1.3 Ordering Layer

Ordering layer uses classical Paxos (by adopting aforementioned optimizations) that one is a well-defined theory of literature. Here, it is unnecessary for explaining it once again. Instead of using the *request_value* from any client, classical Paxos achieves consensus on the *request_id* available in the *stable-ids*. Every Learner learns *request_id* sequentially as per the instance numbers of classical Paxos and inserts these *request_id* into the *decided* set.

When the leader fails, only sequencers are required to participate in leader election process, one of the non-faulty sequencer assumes the role of the leader. Clients, disseminators and learners are not required to know who one is the leader. In HT-Paxos, leader election process does not affect *request* dissemination (i.e., no burden on disseminator and learner sites, unlike S-Paxos).

In HT-Paxos, all sequencers maintain only two sets (i) *stable_ids* set and (ii) *decided* set (Unlike S-Paxos, where every replica maintains four sets). Leader sequentially proposes a *request_id* from the *stable_ids* set, in a new instance of classical Paxos (up to the allowable number of instances at a time). When leader learns a *request_id* after receiving phase 2b messages (of classical Paxos), it inserts this *request_id* into the *decided* set and then deletes this *request id* from the *stable_ids* set. New leader always make it sure that before proposing new *request_id* from *stable_ids,* all the *request_ids* received in phase 1b messages (of classical Paxos) must be decided by as usual working of classical Paxos.

Unlike S-Paxos, HT-Paxos does not require *proposed* and *reproposed* sets. Even though, same optimization (i.e., no duplicate *request_id* will be proposed by the new leader) as claimed in S-Paxos will be achieved here.

### 4.2 Optimizations of HT-Paxos

Before multicasting any *request* to all disseminators and learners, a disseminator can wait for a certain time for more *requests* from one or more clients, and then group them into a *batch*, assign them a unique *batch_id*, after that, this disseminator multicasts < *batch_id*, *batch* > message to all disseminators and learners. Upon receiving a < *batch_id*, *batch* > message any disseminator replies < *batch_id* > message to that disseminator and multicasts < *batch_id* > to all sequencers. Rest of the procedure of HT-Paxos applies to the *batch_id,* similarly as the *request_id*. Such as, at sequencers, *stable_ids* set will contain *batch_ids* and classical Paxos will order the *batch_ids*.

Since the ordering layer uses the classical Paxos, it can use the traditional optimizations of batching and pipelining, as well as any other optimization that applies to the classical Paxos.

Another optimization is to piggyback the acknowledgments on the messages used to forward batches, disseminator sends separate acknowledgment messages only in the absence of such messages. This optimization is especially effective when the system is under high load.

In our protocol, we have used two LANs. However, we may use one or more LANs depending upon various factors. These factors may be either technological or economical. Use of multiple LANs may increase the reliability and performance of the communication network. As per [19], increasing more multicast group can degrade the performance of the communication network. We can reduce multicast groups per LAN using more LANs. This may have a positive impact on performance. If we do not have a technology for required bandwidth in a LAN then in such case, use of multiple LANs can provide the required bandwidth using same technology.

Further optimization that we are proposing increases the fault tolerance of the system for any given number of total computing nodes. In this optimization, we can assume that all disseminator sites also have a sequencer. This optimization may increase fault tolerance of the system but at the cost of comparatively (as compare to HT-Paxos without this optimization) lower throughput of the system. However, throughput under this optimization is still better than any other aforementioned high throughput protocols. We believe that increasing too much fault tolerance at the cost of performance is unnecessary for any large clustered data center, since, massive failures are the rarest events.

## 4.3 Safety

### 4.3.1 Safety Criteria

For the safety of any protocol that implements state machine replication, no two learners can learn the values in different order despite any number of (in our case, non-Byzantine) failures.

### 4.3.2 Proof of Safety (Sketch)

Our proposed protocol fulfills the safety requirement by adopting the following provisions,

*Nontriviality:* learners can learn only values (*client requests or batches)* as indicated by classical Paxos.

*Nontriviality* ensures that learners can learn only the proposed values (*client requests*). As per the proposed protocol, leader of the classical Paxos can only propose the *request_id* or *batch_id* that corresponds to client *requests*; therefore, learners can learn only the *request_id* or b*atch_id* and hence, corresponding *request* or *batch of requests*.

*Consistency:* learners can learn the requests only in same sequence as indicated by classical Paxos.

Since, classical Paxos is a well-proven theory of literature that guarantees safety; therefore, no two learners can learn the values (*client requests*) in different order. ∎

## 4.4 Progress

HT-Paxos ensures that if any client receives a reply for their *request* or *request* becomes an element of *stable_id* set at any disseminator then all available learners will surely learn that *request*. Moreover, protocol also ensures that if client does not crash for an enough time then client will definitely receive a reply for their *request*.

### 4.4.1 Requirements for ensuring progress

At least $\lfloor m/2 \rfloor +1$ sequencers out of total *m*, $\lfloor n/2 \rfloor +1$ disseminators out of total *n* and one learner are always required to remain non-faulty for the progress of the proposed protocol (these requirements are only for ensuring progress, safety does not require these conditions).

*4.4.2 Proof of Progress (sketch)*

As per the protocol, if any client sends a *request* to any disseminator, there could be two cases disseminator may be faulty or non-faulty, if disseminator is faulty then client will not receive a reply for this *request*, therefore, client will resend the request to any randomly chosen disseminator. Since system always has at least a majority of non-faulty disseminators, therefore, there is a fair chance that one of the non-faulty disseminator will receive the client *request*.

If non-faulty disseminator receives a request from any client, after that this disseminator may or may not fail before forwarding the *request* to all disseminators and learners. If disseminator fails then client will not receive a reply, therefore, will resend the *request* to any randomly chosen disseminator. This phenomenon may repeat up to maximum *f* times, where, $f = \lfloor n/2 \rfloor$, because system may have only maximum *f* faulty disseminators.

If disseminator does not fail and forwards the *request* to all disseminators and learners then some or all disseminators and learners may or may not receive *request* due to the message loss. If no disseminators receive *request* due to message loss and sender disseminator fails then client will not receive a reply, in this case, client will resend the *request*. This phenomenon may repeat up to maximum *f* times, because system may have only maximum *f* faulty disseminators.

If some or all disseminators receive *request* from any disseminator, then all such disseminators reply < *request_id* > message. If disseminator does not receive replies from at least a majority of disseminators in a certain time limit for the *request*, in addition, on observing that *request_id* is not an element of *decision* set, then it multicasts *request_id* to all disseminators. If *request_id* is an element of *decision* set, it means, at least (*f* + 1) disseminators have the *request*. Since, ordering layer can decide *request_id* only when it is an element of *stable_ids* set. *Request_id* can become an element of *stable_ids* set only when at least (*f* +1) disseminators have the *request*.

Moreover, on receiving a *request_id* by any disseminator, such that corresponding *request* is not available at this disseminator, then this disseminator sends a < Resend, *request_id* > message to a disseminator from where it has received *request_id*. In the reply of this message, disseminator receives the corresponding *request*. Furthermore, if disseminator observes that the *request_id* is an element of *decision* set then periodically sends a < Resend,

*request_id* > message to any other disseminator. If learner is not on the disseminator's node, then on learning a *request_id*, if corresponding *request* is not available then it periodically sends < Resend, *request_id* > message to any disseminator until it receives the *request*.

Statements in the above two paragraphs ensure that all non-faulty disseminators and learners will receive the *request*. Non-faulty disseminator that received the client *request* either receives a majority of reply messages or observes that *request_id* is an element of *decided* set. Hence, client will receive a reply if it does not fail, because disseminator will periodically send reply to the client until it receives a reply or detects a failure.

Leader will receive same < *request_id* > messages from at least $(f+1)$ disseminators, because at least $(f+1)$ disseminators are always non-faulty as per our assumption, all non-faulty disseminators have the *request* as per the above paragraph and all disseminators that have *request* periodically multicasts < *request_id* > to all sequencers. Hence, *request_id* will definitely become an element of *stable_id* set at the leader and then classical Paxos will order all the elements of *stable_id* set. As we already know, classical Paxos guarantees progress under aforementioned requirements, therefore, at least one non-faulty learner will definitely learn the *request_id*. Since, all non-faulty learners have the corresponding *request* as per statements of above paragraph. Therefore, all non-faulty learners will learn the corresponding *request*.

Hence, we can say that under aforementioned specific requirements, HT-Paxos ensures progress.

## 5. Comparative Analysis

As we are aware, the workload of most of the real time applications that use state machine replication for increasing availability is increasing day by day. Therefore, requirement of high throughput is also increasing accordingly. We can increase throughput by increasing the processing power of computers and increasing the bandwidth of communication network. Every time this solution for higher throughput may not be practical for either technological and/or economical reasons. Because, replacement of existing computers and communication network with higher processing power computers and higher bandwidth communication network may be a costly affair and may not be practical every time. Moreover, there may be the case that higher technology of computers and communication network may not be available every time.

Alternatively, we can adopt a more scalable and throughput efficient protocol, i.e., a protocol that requires comparatively less computation at individual computers and less traffic at individual LANs. In addition, it may increase throughput by increasing more computers and more LANs. Although after a certain limit, we cannot scale up the system because of coordination overload, instead, it may start reducing the throughput after certain limit. This limit depends on the protocol that we use.

Earlier versions of Paxos (like classical Paxos, fast Paxos or generalized Paxos) lacks in terms of scalability and throughput, because, particularly leader has more processing and bandwidth requirements. Other variants of Paxos like ring Paxos, multi-ring Paxos and S-Paxos increase the scalability and throughput by reducing the processing and bandwidth requirements, especially at the leader. We are going to compare the processing and bandwidth requirements among various Paxos protocols that affects system scalability and throughput.

## 5.1 Processing requirements

In general, for state-machine replication protocols, processing requirements at any individual computer reduce, if computer requires to response or process a less number of messages. Therefore, we require analyzing the number of incoming and outgoing messages. For the analysis, we are considering here the case of normal operations.

Moreover, we also require some processing for the transmission of the data. If any individual computing node requires higher data transmission then it requires higher processing requirement for the data transmission. We will discuss this requirement in the next *bandwidth requirements* section.

### *5.1.1   Message Analysis of HT-Paxos*

Let, we assume various clients issue total number of *n requests* per unit time and total *m* disseminators are there then on an average, each disseminator receives *n/m requests* per unit time. Moreover, we assume each disseminator makes a batch of *n/m requests* per unit time. We further assume that the leader makes a batch of *m batch_ids* and total *s* sequencers are there.

For the processing of client *requests* of one unit time, we are analyzing now the required number of messages.

#### *5.1.1.1   At any disseminator site*

Total incoming messages $=\left(\left(n/m\right)+2m\right)$

Since, disseminator will receive *m/n requests* directly from the clients, *m* batches from all disseminators (including self), *m* reply messages *<batch_id>* from all disseminators (including self) and one decision message containing *m batch_ids* from the leader (since, learner is also on disseminator' site).

Total outgoing messages $= (m+3)$

Since, one multicast of their own batch to all disseminators and learners, per batch one reply <batch_id> message, one multicast < batch_id > message to all sequencers and a reply message to the client.

Total messages at a disseminator's site $= (3m+(n/m)+3)$

### 5.1.1.2 At the leader site

Total incoming messages $= (m+\lfloor s/2 \rfloor)$

Since, leader receives $m$ batch_ids and $\lfloor s/2 \rfloor$ phase 2b messages of classical Paxos, as leader is also a one of the acceptor of classical Paxos, so $\lfloor s/2 \rfloor +1$ sequences (acceptors of classical Paxos) create a required majority.

Total outgoing messages $= 2$

Since, leader multicasts one phase 2a message to majority of sequencers (acceptors of classical Paxos), multicasts a decision message to all sequencers, disseminators and learners.

Total messages at the leader's site $= (m+\lfloor s/2 \rfloor +2)$

### 5.1.1.3 At any sequencer site (other than the leader)

Total incoming messages $= m+2$

Since, sequencer receives $m$ batch_ids, one phase 2a message of classical Paxos and one a decision message from the leader.

Total outgoing messages $= 1$

Since, sequencer only sends a phase 2b message of the classical Paxos.

Total messages at a sequencer $= m+3$

### 5.1.1.4 At any learner site (without disseminator)

Total incoming messages $= m+1$

Since, learner receives $m$ batches and one decision message from the leader.

In normal operations no outgoing message.

Therefore, Total messages $= m+1$

*5.1.2    Message analysis of Ring Paxos*

Because of the processing, bottleneck may be at the leader. Therefore, we are calculating here the total number of messages at the leader. Just like HT-Paxos, we assume that out of total *n requests* leader makes *m* batches of *n/m requests* each.

Total incoming messages $= n + m$

Since, leader will receive *n requests* directly from the clients and for *m* batches leader will receive *m* messages from the last acceptor of the ring.

Total outgoing messages $= n + m + 1$

Since, leader will send *n* reply messages to the clients, for *m* batches leader will ip-multicast *m* messages to all acceptors and learners and ip-multicast one decision message containing *m batch_ids* to all acceptors and learners.

Total messages at the leader's site $= 2(n + m) + 1$

*5.1.3    Message analysis of S-Paxos*

Because of the processing, bottleneck may be at the leader. Therefore, we are calculating here the total number of messages at the leader. Just like HT-Paxos, we assume that various clients issue total number of *n requests* per unit time and total *m* disseminators are there then on an average, each disseminator receives *n/m requests* per unit time. Moreover, we assume each disseminator makes a batch of *n/m requests* per unit time. We further assume that the leader makes a batch of *m batch_ids*.

Total incoming messages $= \begin{pmatrix} \left( (n/m) + m + m^2 \right) \\ + \lfloor m/2 \rfloor + 1 \end{pmatrix}$

Since, leader will receive *n/m requests* directly from the clients, *m* batches from all disseminators (including self), per batch *m* reply messages *<batch_id>* from all disseminators (including self), $\lfloor m/2 \rfloor$ messages of phase 2b of classical Paxos and one decision message from self.

Total outgoing messages $= n/m + m + 3$

Since, *n/m* reply messages to the clients, per batch one multicast of reply *<batch_id>* message to all replicas, one multicast of their own batch to all replicas, one multicast of phase 2a message of classical Paxos and multicast of one decision message to all replicas.

Total messages at the leader's site = $\begin{pmatrix} m^2 + 2(n/m) \\ +2m + \lfloor m/2 \rfloor + 4 \end{pmatrix}$

### 5.1.4 Message analysis of classical-Paxos

Because of the processing, bottleneck may be at the leader. Therefore, we are calculating here the total number of messages at the leader under batching optimization for reducing the number of messages. Just like HT-Paxos, we assume that out of total *n requests* leader makes *m* batches of *n/m requests* each. Let, there are total *m* acceptors.

Total incoming messages = $n + m * \lfloor m/2 \rfloor$

Since, *n* client requests, per batch $\lfloor m/2 \rfloor$ messages of phase 2b.

Total outgoing messages = $n + 2m$

Since, *n* reply messages to the clients, per batch one multicast of phase 2a and one multicast of decision message.

Total messages at the leader's site = $2(n+m) + m * \lfloor m/2 \rfloor$

### 5.1.5 Comparative message analysis

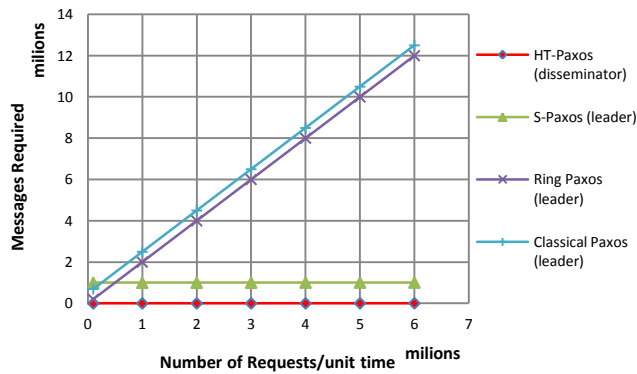

**Fig. 1** Comparison among mentioned variants of Paxos for the messages requirements at the busiest computing nodes, where *m* = 1000, *s* = 20

As we can see in Fig. 1, high number of messages in classical Paxos and in ring Paxos are because of all client communications are through the leader. S-Paxos and HT-Paxos decentralize the client communication i.e., clients may approach any disseminator. Message advantage of HT-Paxos over S-Paxos is because of the fact that in S-Paxos, every disseminator is required to reply to every other disseminator, in our HT-Paxos reply goes to only one disseminator and disseminator sites are not concerned with the most of the messages of ordering layer.

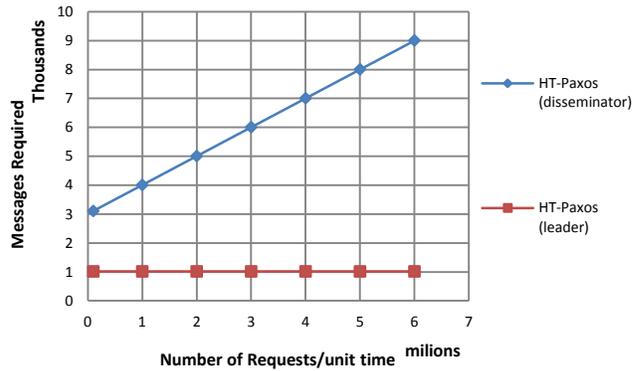

**Fig. 2** Comparison between any one disseminator and the leader of the HT-Paxos for the messages requirements, where *m* = 1000 and *s* = 20

We can see in Fig. 2 that leader in HT-Paxos is very much lightweight as compared to any disseminators. It means bottleneck may not be at the leader's site in HT-Paxos (if optimized for throughput rather than fault tolerance).

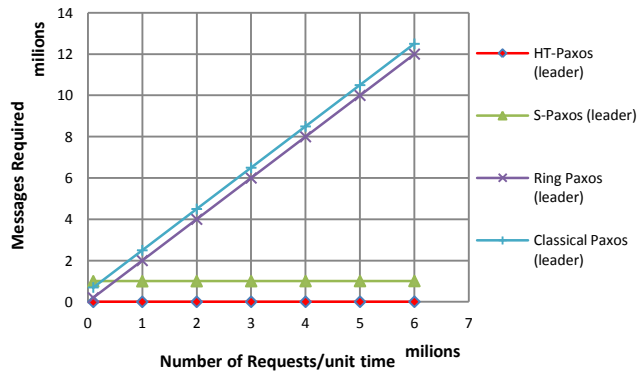

**Fig. 3** Comparison among mentioned variants of Paxos for the messages requirements at the busiest computing nodes, where m=1000 and every disseminator site also has a sequencer and a learner, i.e., fault tolerant version of HT-Paxos

In fault tolerant version of HT-Paxos, ordering layer messages also become the part of the busiest computing node (leader's site) as similar to the S-Paxos. The message advantage of this version of HT-Paxos over S-Paxos is because of the aforementioned reply mechanism of disseminators.

## 5.2 Bandwidth requirements

Size of data and number of messages required to transmit by any computer affects the bandwidth requirements of the communication network. If any protocol requires more number of messages than due to message overhead, more data will pass through the communication network, hence requires higher bandwidth. Bottleneck may be the bandwidth of communication network due to large data size and high number of messages. In any data centre, if bandwidth is bottleneck, then there are two options either replace the lower bandwidth LAN with higher bandwidth LAN, or adopt multiple LANs of same bandwidth. First option may not be practical for either technological or economical reasons. In data centers we do not requires big cables, therefore, it is not a costly affair, hence not a big issue in any large data centre.

However, if any computing node requires transmitting and receiving more data, then bottleneck may be the network sub system of computing node that works for the transmitting and receiving of the data. Replacements of computing nodes with higher processing powers may really be a big issue, because it may be a costly affair.

Therefore, we are checking the bandwidth requirements of individual computing nodes of the various variant of Paxos. For that, we are considering the same assumptions as in the previous section. Moreover, we are assuming that message overhead 64 bytes (as ip packet header, Ethernet frame preamble, header, footer, gap and other network protocols like ARP etc create overheads. Bigger message overhead will be in the favor of our protocol, because it requires less messages as mentioned above), and *request_id*, *batch_id*, *round number*, *instance number* are 4 bytes each.

What incoming and outgoing messages are there, on that basis we can calculate the incoming and outgoing data per unit time.

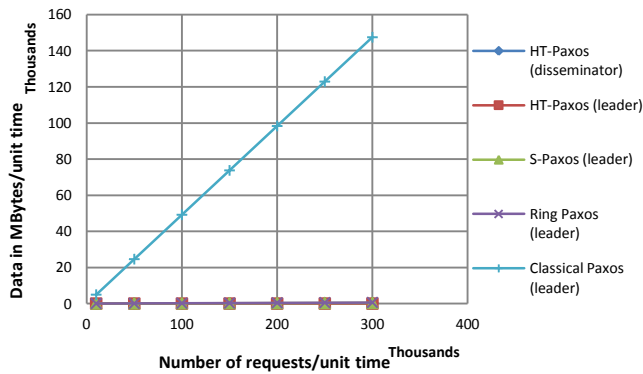

**Fig. 4** Comparison of bandwidth requirements at the mentioned computing nodes of the various mentioned variant of Paxos, where $m = 1000$, $s = 20$ and data size of request = 1k bytes.

In any clustered data center, if we use classical Paxos then leader of classical Paxos handles extremely large amount of data (as mentioned in Fig. 4) just because protocol achieves consensus on *request* (or *batch*) instead of *request_id* (or *batch_id*). Other variants of Paxos for high throughput achieves consensus on *request_id* (or *batch_id*) instated of *request* (or *batch*) because, in general, *request_id* (or *batch_id*) remains very small as compare to the corresponding *request* (or *batch*).

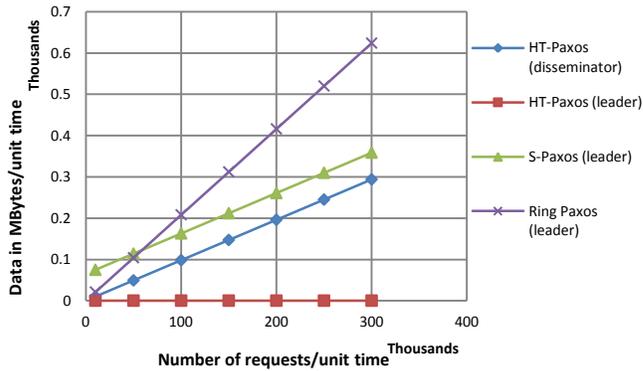

**Fig. 5** Comparison of bandwidth requirements at the mentioned computing nodes of the various mentioned variant of Paxos, where $m = 1000$, $s = 20$ and data size of request = 1k bytes.

If number of *requests* increases, the leader of ring Paxos handles large amount of data as compare to other high throughput Paxos (as shown in Fig. 5). Major reason is that the leader handles all client communications. Moreover, in case of fewer *requests*, ring Paxos performs better than S-Paxos, major reason for this is the comparatively large number of reply messages at the disseminators. Furthermore, disseminator of HT-Paxos handles less data because of decentralized client communications like S-Paxos; in addition, it reduces the number of reply messages at the disseminators. Furthermore, leader of HT-Paxos is significantly lightweight because it handles lightweight *request_ids* or *batch_ids.*

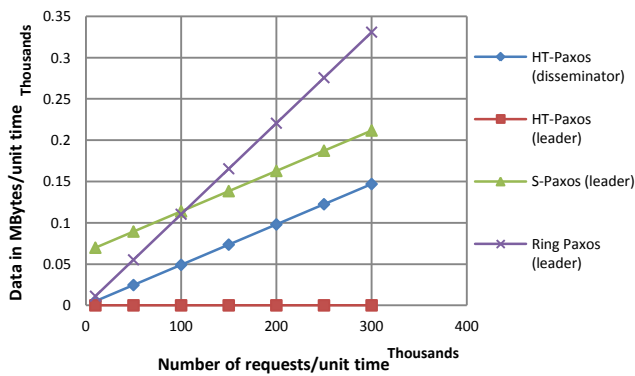

**Fig. 6** Comparison of bandwidth requirements at the mentioned computing nodes of the various mentioned variant of Paxos, where $m = 1000$, $s = 20$ and data size of request = 512 bytes.

As the data size of the client *request* reduces, we can observer that the gap of S-Paxos with HT-Paxos widens as shown in Fig. 6, this is because of high ratio of metadata in S-Paxos as compare to HT-Paxos. Moreover, S-Paxos becomes better than ring Paxos in such case only after more number of requests/per unit time.

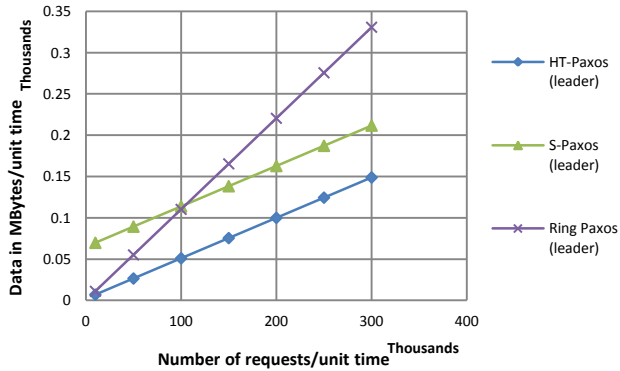

**Fig. 7** Comparison among mentioned variants of Paxos for the messages requirements at the busiest computing nodes, where m=1000, every disseminator site also has a sequencer and a learner, i.e., fault tolerant version of HT-Paxos and data size of request = 512 bytes.

In fault tolerant version of HT-Paxos, leader's site combines the dissemination and ordering layer data, but ordering layer data is too low therefore, impact in data terms at leader's site is not too much as shown in Fig. 7.

### 5.3 Latency

HT-Paxos and S-Paxos both take six message delays for learning the client requests in the best case (as we are considering the message-optimized version of classical Paxos in the ordering layer). Classical Paxos takes four message delays in message-optimized version and three message delays otherwise in the best case. Moreover, fast Paxos and generalized Paxos take only two message delays in the best case. While, Ring Paxos take ($m$ +2) message delays in the best case, where, $m$ represents the total number of acceptors in the ring.

### 5.4 Response time

HT-Paxos takes only four message delays for responding to the client *request* in the best case, because we have chosen a slightly optimistic approach for sending a reply to the client i.e. on being sure that the *request* is available at any majority of disseminators, disseminator who has received the *request* from the client sends the corresponding reply. Since, under mentioned assumptions, *request* will definitely execute. However, if clients want to get a reply only after the execution of *requests,* as in the case of S-Paxos then HT-Paxos will also take six message delays like S-Paxos. While, Ring Paxos takes ($m$ +2) message delays in the best case, where, $m$ represents the total number of acceptors in the ring. Moreover, in this regard classical Paxos has a comparatively good performance as it takes only four message delays.

## 5.5 Other Related Work

Zab [24] is a variant of the Paxos designed for the primary-backup data replication systems such as yahoo's Zookeeper coordination service. In zookeeper, any client approaches to any replica (either leader or follower) for their *requests*. Follower replica forwards all update *requests* to the primary replica for taking the services of state-machine replication protocol Zab. Zab is a centralized protocol that has one primary that disseminates the update *requests* to all other replicas and the leader that generally is on the same primary site works for ensuring a proper order. However, HT-Paxos is a more decentralized state-machine replication protocol that has multiple disseminators, any client for their update *request* may directly approach any replica that has a disseminator, and after that disseminator forwards the update *request* to all other replicas. In case of read *request* client may approach any replica. Because of the centralized nature of the Zab, bottleneck may be the resources of the leader's site (or primacy's site as Zab considers both on the same site) in any large clustered data centers. Therefore, throughput and scalability will obviously be less in any large clustered data center where workload is very high.

Mencius [18] takes an alternative approach that is a moving sequencer approach [9] to prevent the leader from becoming the bottleneck. Mencius partitions the sequence of consensus protocol instances among all replicas and each replica becomes a (initial) leader of an instance in a round-robin fashion. Protocol excludes all failed replicas by adopting a reconfiguration mechanism. This protocol is a quite decentralized protocol like HT-Paxos. However, every replica failure requires a reconfiguration of the system this is not the case of HT-Paxos. Moreover, even in the case of failure free execution, leader of Mencius does the work of dissemination as well as ordering. However, in throughput-optimized version of HT-Paxos, leader is only responsible for ordering of *request_ids* and is very much lightweight. Under a large clustered data center and heavy load environment that is the basic motivation of this paper, leader of Mencius will handle more number of messages as well as more data as compared to any disseminator or the leader. Performance of Mencius against fault-tolerant version of HT-Paxos in failure free environment may be quite comparable. However, design goal of Mencius was to provide an optimized state-machine replication protocol for WAN environment. Contrary to this HT-Paxos is for clustered environment.

LCR [22] is a high throughput state-machine replication protocol base on virtual synchrony model [20] of data replication instead of Paxos. LCR arranges replicas along a logical ring and uses vector clocks for message ordering.

LCR is a high-throughput protocol, where all replicas are equally loaded, thereby utilizing all available system resources. However, latency and response time increases linearly with the number of processes in the ring. For any large clustered data center, this will be very significant. Although LCR has slightly better bandwidth efficiency, Furthermore, in LCR, every failure of the replica requires a view change for ensuring progress and *perfect failure detection* is required i.e. erroneously considering a process to have crashed is not tolerated, it implies stronger synchrony assumptions.

State partitioning [8] is another technique that can achieve scalability. Multi-Ring Paxos [27] uses this concept and keeps various logical groups. Each logical group has an instance of ring Paxos (in optimized version, multiple logical groups may also have a single instance of ring Paxos). Any learner may subscribe to any one or more logical groups. If a learner subscribes to multiple groups then it uses a deterministic procedure to merge messages coming from different instances of ring Paxos. However, HT-Paxos can easily adopt the concept of state partitioning by slightly changing the dissemination layer, as disseminator can multicast the *request* to only interested learners, while ordering layer would deliver the order to all learners (like S-Paxos). In ring Paxos or in Multi-Ring Paxos, any failure of acceptor requires a view change, Moreover, latency and response time increases linearly with the number of acceptors in the ring.

## 6. Conclusion And Future Work

HT-Paxos is a variant of Paxos designed for large clustered data centers that achieves significantly high throughput and scalability. It achieves all this by further offloading the leader i.e. HT-Paxos is very much decentralize protocol. As we are aware, the primary focus of earlier versions of Paxos was fault tolerance and latency, because on that time throughput requirement was comparatively very low. In Paxos based protocols, the major obstacle for high throughput was bottleneck at the leader. In such systems, very soon on increasing more computing nodes fault tolerance increases rather than throughput. Practically this is highly undesirable, because massive failures could be a very rare event in the clustered data centers. Instead, it is quite more desirable in the data centers that on increasing more computing nodes, it should increase performance in terms of throughput.

Moreover, throughput may be limited because of processing power of CPU or data handling capacity of network sub system of any computing node or bandwidth of communication networks. As commuting resources are generally very much costly as compare to data cables, because in clustered data centers length of data cables required may not be too much as compared to WAN environment. Therefore, high throughput state-machine replication protocols should avoid bottleneck of CPU and network subsystems at any computing node through less computing requirements of CPU and less bandwidth requirements at any individual computing node. Proposed HT-Paxos achieves all these goals very significantly for improvement of throughput and scalability. Furthermore, on the same time, latency and response time of the HT-Paxos as compare to other high throughput state-machine replication protocols is quite less.

As future work, we plan to apply our technique to Byzantine faults, and will optimize HT-Paxos for WAN.